\documentclass[12pt]{article}

\usepackage{graphicx}
\usepackage{amsmath}
\usepackage{amssymb}
\allowdisplaybreaks

\begin{document}

\baselineskip=17.5pt plus 0.2pt minus 0.1pt

\renewcommand{\theequation}{\thesection.\arabic{equation}}
\renewcommand{\thefootnote}{\fnsymbol{footnote}}
\makeatletter
\@addtoreset{equation}{section}
\def\CR{\nonumber \\}
\def\pt{\partial}
\def\be{\begin{equation}}
\def\ee{\end{equation}}
\def\bea{\begin{eqnarray}}
\def\eea{\end{eqnarray}}
\def\eq#1{(\ref{#1})}
\def\la{\langle}
\def\ra{\rangle}
\def\hyp{\hbox{-}}

\begin{titlepage}
\title{
\hfill\parbox{4cm}
{ \normalsize YITP-04-01 \\{\tt hep-th/0401079}}\\
\vspace{1cm}Field theory on evolving fuzzy two-sphere}
\author{
Naoki {\sc Sasakura}\thanks{\tt sasakura@yukawa.kyoto-u.ac.jp}
\\[15pt]
{\it Yukawa Institute for Theoretical Physics, Kyoto University,}\\
{\it Kyoto 606-8502, Japan}}
\date{\normalsize May, 2004}
\maketitle
\thispagestyle{empty}

\begin{abstract}
\normalsize
I construct field theory on an evolving fuzzy two-sphere, which is 
based on the idea of evolving non-commutative worlds of the previous paper \cite{Sasakura:2003ke}.
The equations of motion are similar to the one that can be obtained by dropping the 
time-derivative term of the equation derived some time ago by Banks, Peskin and Susskind for
pure-into-mixed-state evolutions. 
The equations do not contain an explicit time, and therefore 
follow the spirit of the Wheeler-de Witt equation. 
The basic properties of field theory such as action, gauge invariance and charge and momentum
conservation are studied. The continuum limit of the scalar field theory shows that the
background geometry of the  corresponding continuum theory is given by 
$ds^2 = -dt^2+t\ d\Omega^2$, which saturates locally the cosmic holographic principle. 
\end{abstract}
\end{titlepage}

\section{Introduction}
Several thought experiments in quantum mechanics and general relativity 
\cite{salecker}-\cite{Adler:1999bu} and in string theory \cite{Yoneya:2000bt} 
show that there exist limits on the measurements of space-time observables such as lengths, 
areas, and positions.
At present the meaning behind of these limits is far beyond our reach, but they suggest 
that our space-time may have some quantum natures, and that the fundamental principle underlying 
string theory may be given by a space-time uncertainty relation \cite{Yoneya:2000bt}. 
To include the effects of such quantum natures of space-time, several authors have proposed various 
algebraic descriptions of quantum space-times by introducing non-commutativity among momentums and coordinates 
\cite{Doplicher:1994zv}, \cite{Snyder:1946qz}-\cite{Kawamura:2003cw}. 

A simple model of a quantum space is a fuzzy two-sphere \cite{Madore:1991bw}. 
This is given by regarding the
three su(2) generators as the coordinates of a three-dimensional space, and the spin of its
representation roughly as the radius of a sphere embedded in the space. 
Since our universe is changing its size, a fundamental interesting question would be how we can 
describe an evolution of a fuzzy sphere to different sizes.\footnote{The splitting process of 
a fuzzy two-sphere was studied in \cite{Balachandran:2003wv}.}
This question becomes rather ambiguous if we do not impose any physical requirements since 
the dimensions of the representation spaces are generally distinct for each representation 
and therefore there exist no unitary maps between them.
Since it is a natural expectation that a perfect sphere should evolve to a perfect one, the SU(2)
symmetry should be respected. In the previous paper \cite{Sasakura:2003ke}, a general method to 
describe such an evolution with symmetry was given based on splitting process of a representation
space to two and tracing out one of them. 
This kind of evolution generally results in pure-into-mixed-state evolutions of a density matrix
or operators.

Pure-into-mixed-state evolutions were proposed by Hawking for quantum field theory to accommodate
quantum gravity \cite{Hawking:1982dj}. His proposal was rewritten in the form of a differential
equation of a density matrix by Banks, Peskin and Susskind (BPS) \cite{Banks:1983by}.
In the next section, it will be shown that the evolution of a fuzzy two-sphere presented 
in the previous paper \cite{Sasakura:2003ke} can be described in a compact form by dropping 
the time-derivative term of the BPS differential equation. 
Then the equation becomes a constraint equation of a density matrix, and does not
contain an explicit time. This absence of an explicit time would be natural from the viewpoint 
of the canonical quantization of general relativity,
since time is a gauge-dependent quantity and the Wheeler-de Witt equation is a constraint 
equation resulting from the gauge-fixing of time.   

To obtain field theory on an evolving fuzzy two-sphere, I will work in the Heisenberg picture
presented in the previous paper \cite{Sasakura:2003ke}. The dynamical variables are fields 
themselves instead of a density matrix like in the original papers of Hawking and BPS. 
I will obtain constraint equations of fields as the equations of motion.
In Section \ref{scalar}, I will give the equations of motion of scalar fields, 
and discuss the actions, gauge invariance and conserved charges and momentums. 
Spinor and gauge fields will be discussed in Section \ref{spinor} and Section \ref{gauge},
respectively.
In Section \ref{continuum}, I will take the continuum limit of the equation of motion of  
a scalar field and obtain the background geometry of the corresponding continuum theory. 
The final section will be devoted to summary and discussions.

\section{Evolution of a fuzzy two-sphere}
\label{twosphere}

To start with the Schwinger's representation of su(2) algebra,
let me consider the following two sets of creation-annihilation operators,
\be
\label{creationannihilation}
[a_i,a^\dagger_j]=\delta_{ij},
\ee
where $i,j=1,2$.
The su(2) generators are given by
\be
L_{k}=\frac{1}{2}a^\dagger_i\sigma^k_{ij}a_j,
\ee
where $\sigma^k_{ij}\ (k=1,2,3)$ are the Pauli matrices, and the repeated indices are summed over. 
These $L_i$ satisfy the su(2) commutation relations,
\be
[L_i,L_j]=i\, \epsilon_{ijk}\, L_k.
\ee
The spin $J=N/2$ representation of su(2) is spanned by 
\be
\left|N,n \right>=\frac{1}{\sqrt{(N-n)!\, n!}} (a^\dagger_1)^{N-n} (a^\dagger_2)^n \left|0\right>,
\ee
where $n=0,\cdots,N$, and the $\left|0\right>$ is the Fock vacuum $a_i\left|0\right>=0$.
As a notation, ${\cal H}_N$ denotes the space spanned by $\left|N,n\right>\ (n=0,1,\cdots,N)$, 
the Hilbert space of the spin $N/2$ representation. 

In this paragraph, let me recapitulate the argument of my previous paper \cite{Sasakura:2003ke}.
Consider splitting process of a state into two,
\be
\label{unitaryevol}
\left|i\right>\rightarrow {C_{i}}^{jk} \left|j\right> \left|k\right>,
\ee
where the repeated indices are summed over. Unitarity and symmetry are assumed to be preserved 
in the process.
Tracing out the second state of the right-hand side, 
the following evolution process is obtained,
\be
\label{pureintomixed}
\left|i\right> \left< i' \right| \rightarrow 
{C_{i}}^{jk} {{C_{i'}}^{j'k'}}^* \eta_{k'k} \left| j\right> \left< j' \right|,
\ee
where $\eta_{k'k}=\left< k'| k \right>$.
Here what evolves can be regarded as a density matrix in the Schrodinger picture or 
operators on the representation space in the Heisenberg picture, respectively.
In the usual quantum mechanics, the two pictures are physically equivalent, but it is not
so in the pure-into-mixed-state evolution \eq{pureintomixed}, as was discussed in 
\cite{Sasakura:2003ke}.  
When the Clebsh-Gordon coefficients are chosen as the ${C_{i}}^{jk}$, an evolution process
of a fuzzy two-sphere can be obtained. 
Choosing the smallest non-vanishing spin $1/2$ for the second state of the right-hand side 
of \eq{unitaryevol}, a monotonous expansion is obtained as 
\bea
\label{evolutionold}
\left|N,n \right>\left<N,n'\right|&\rightarrow& \frac{\sqrt{(N-n+1)(N-n'+1)}}{N+2}
\left|N+1,n\right>\left<N+1,n'\right| \cr 
&&\ \ \ +\frac{\sqrt{(n+1)(n'+1)}}{N+2}\left|N+1,n+1\right>\left<N+1,n'+1\right|.
\eea

The same evolution can be derived from splitting process of fuzzy two-spheres discussed
in \cite{Balachandran:2003wv}. Their discussions are based on the observation that
the coproduct ${\cal A}\rightarrow {\cal A}\otimes {\cal A}$ of a Hopf algebra may be  
regarded as splitting process of a fuzzy space defined by ${\cal A}$. For the splitting
process of a spin $J/2$ fuzzy two-sphere to those with spin $K/2$ and $L/2$,
they obtained a map $\Delta_{J,KL}$ defined by
\be
\label{spilitting}
\Delta_{J,KL}(\left|J,j\right>\left<J,j'\right|)
=\sum_{k,k',l,l'}C(K,L,J;k,l)C(K,L,J;k',l')
\left|K,k\right>\left<K,k'\right| \otimes \left|L,l\right>\left<L,l'\right|,
\ee
where $C(K,L,J;k,l)$ are the Clebsh-Gordon coefficients.
The map $\Delta_{J,KL}$ has the properties,
\bea
\Delta_{J,KL}(A^\dagger)&=&\Delta_{J,KL}(A)^\dagger, \cr
{\rm Tr}_J (A)&=&{\rm Tr}_{KL} (\Delta_{J,KL}(A)),\cr
\Delta_{J,KL}(A)\Delta_{J,KL}(B)&=&\Delta_{J,KL}(AB),
\eea
where $A,B$ are arbitrary operators on the spin $J/2$ representation space. 
These properties assure the unitarity of the splitting process. By taking $J=N, K=N+1, L=1$ 
at \eq{spilitting} and 
tracing out $\left|L,l\right>\left<L,l'\right|$, one obtains a map,
\bea
\label{evolutionnew}
\Delta'_{N, N+1}(\left|N,n \right>\left<N,n'\right|)&=& \frac{\sqrt{(N-n+1)(N-n'+1)}}{N+2}
\left|N+1,n\right>\left<N+1,n'\right| \cr
&& +\frac{\sqrt{(n+1)(n'+1)}}{N+2}\left|N+1,n+1\right>\left<N+1,n'+1\right|,
\eea
which is the same as \eq{evolutionold}. Therefore the pure-into-mixed state evolution 
\eq{evolutionold} can be physically interpreted as an evolution of 
a `main' fuzzy space under a unitary process of emitting a `baby' fuzzy space, 
as was previously discussed in \cite{Sasakura:2003ke}. 
 
The evolution \eq{evolutionold} can be compactly expressed in terms of 
the creation-annihilation operators introduced in \eq{creationannihilation}.
In fact, the right-hand side of \eq{evolutionold} equals 
\be
\frac1{N+2} \sum_{i=1}^2 \, a^\dagger_i \left|N,n \right>\left<N,n'\right| a_i.
\ee
Therefore the evolution \eq{evolutionold} can be rewritten as
\be
\label{eachevol}
(\hat{N}+1){\cal O}_{N+1}=\sum_{i=1}^{2} a^\dagger_i \, {\cal O}_N \, a_i,
\ee
where $\hat{N}$ is the number operator for the creation-annihilation operators,
\be
\hat{N}=\sum_{i=1}^2 a^\dagger_i a_i,
\ee 
and ${\cal O}_N$ denotes a density matrix on ${\cal H}_N$ in the Schrodinger picture 
or an operator on ${\cal H}_N$ in the Heisenberg picture, respectively. 
It is clear that the su(2) symmetry is preserved in \eq{eachevol}. Let us define 
\be
{\cal O}=\sum_{N=N_0}^\infty {\cal O}_N,
\ee
where $N_0$ represents the initial boundary.
Then \eq{eachevol} can be rewritten as a constraint equation of ${\cal O}$,
\be
\label{constraintsu2}
(\hat{N}+1){\cal O}-\sum_{i=1}^{2} a^\dagger_i \, {\cal O} \, a_i=(N_0+1){\cal O}_{N_0},
\ee
where the right-hand side gives an initial boundary condition.
 
Banks, Peskin and Susskind \cite{Banks:1983by} 
described the evolution of pure states into mixed states by
the following differential equation for a density matrix $\rho$,
\be
\label{BPS}
\dot{\rho}=-i\,[H_0,\rho]-\frac12\sum_{\alpha\beta} h_{\alpha\beta}(Q^\beta Q^\alpha \rho
+\rho\, Q^\beta Q^\alpha-2 Q^\alpha \rho \, Q^\beta),
\ee
where $h_{\alpha\beta}$ is Hermitian, and $H_0$ and $Q^\alpha$ are Hermitian operators. 
With $H_0=0$ and an appropriate choice of $h_{\alpha\beta}$ and $Q^\alpha$, 
the right-hand side of \eq{BPS} agrees with the left-hand side of the constraint equation 
\eq{constraintsu2} up to $constant \times {\cal O}$. 
This constant shift merely changes the normalization of each ${\cal O}_N$. 
The normalization is crucial for a density matrix, but is not necessarily 
in the Heisenberg picture, which will be used in the following sections. 
Thus the constraint equation \eq{constraintsu2} essentially agrees 
with the BPS equation \eq{BPS} after dropping the term of time-derivative. 

The absence of time in \eq{constraintsu2} is natural from the 
viewpoint of the canonical quantization of general relativity.  
Time is a gauge-dependent quantity in general relativity,
and its gauge-fixing leads to a constraint, 
the Wheeler de-Witt equation. Therefore the constraint \eq{constraintsu2} follows the 
spirit of the Wheeler de-Witt equation. 
Time will be recovered by identifying it with a physical observable. In Section \ref{continuum},
where the continuum limit will be taken, the size of the sphere will play the role of time.

\section{Scalar field}
\label{scalar}
In this section, I will discuss the equations of motion of scalar fields on the evolving 
fuzzy two-sphere of the previous section. I will construct them by analogy of the continuum theory.  
The correct correspondence to the continuum theory will be discussed later 
in Section \ref{continuum}.
I will also study the actions, gauge invariance, current conservation,
and conserved charges and momentums. 
In this and the following sections, the Heisenberg picture will be utilized, 
where dynamical variables are fields themselves instead of a density matrix in the Schr\"odinger 
picture. As was discussed in \cite{Sasakura:2003ke}, the two pictures are
physically inequivalent. Concerning the conserved quantities, the choice of the Heisenberg
picture is crucial. This point will be further discussed in Section \ref{discussions}. 

Let me assume that a scalar field $\phi$ is an Hermitian operator given by
$\phi=\sum_N \phi_N$, where each $\phi_N$ is an Hermitian operator on each ${\cal H}_N$. 
The equation of motion of a massless scalar field $\phi_c$ 
in the continuum has the following form,
\be
\left(-\nabla^2+\left(\frac{\pt}{\pt t}\right)^2 \right) \phi_c =0.
\ee
The second order time-derivative may be replaced by a finite difference,
\be
\left(\frac{\pt}{\pt t}\right)^2 \phi_c(t) \sim \frac{
 \phi_c(t+\Delta t)+\phi_c(t-\Delta t)-2\phi_c(t)}{(\Delta t)^2},
\ee
if $\Delta t$ is small enough.
Since the shift of time may be regarded as the shift of the representation in the present model, 
\bea
\phi_c(t+\Delta t)&\sim& \sum_{i=1}^{2} a^\dagger_i \, \phi \, a_i,\cr
\phi_c(t-\Delta t)&\sim& \sum_{i=1}^{2} a_i \, \phi \, a^\dagger_i. 
\eea
The spatial part may be replaced by \cite{Madore:1991bw}
\be
\nabla^2 \phi_c \, \sim\, -\sum_{i=1}^3 [L_i,[L_i,\phi]].
\ee
Therefore I propose the following equation of motion for a massless Hermitian scalar field
on the evolving fuzzy two-sphere,
\be
\label{eomscalarone}
\sum_{i=1}^3 [L_i,[L_i,\phi]]+\sum_{i=1}^{2} a^\dagger_i \,  \phi \, a_i 
+ \sum_{i=1}^{2} a_i \, \phi \, a^\dagger_i - (2 \hat{N}+2) \phi =0.
\ee
Here the last term has been determined from the requirement that 
the massless equation of motion should have the trivial solution of constant, 
$\phi=\sum_N {1}_N$.
It is easy to check that \eq{eomscalarone} can be derived from the BPS equation \eq{BPS},
by setting 
\bea
\dot{\rho}&=&H_0=0, \cr
Q^\alpha&=&\left( L_i,\frac{a_i+a^\dagger_i}{2},\frac{i(a_i-a^\dagger_i)}{2}\right),
\eea 
with a diagonal metric, $h_{\alpha\beta}=(1,1,1,-1,-1,-1,-1)$. 

The equation of motion \eq{eomscalarone} can be also rewritten as 
\be
\label{eommassless}
\sum_{i=1}^3 [L_i,[L_i,\phi]]-\sum_{i=1}^{2} [a^\dagger_i,[a_i,\phi]]=0.
\ee 
In this expression the existence of the constant solution, $\phi=\sum_N {1}_N$, is 
clearly seen. From the discussions in Section \ref{continuum}, it will turn
out that the equation of motion with a potential term is given by  
\be
\label{eompotential}
\sum_{i=1}^3 [L_i,[L_i,\phi]]-\sum_{i=1}^{2} [a^\dagger_i,[a_i,\phi]]
+(\hat{N}+1) V'(\phi)=0,
\ee
where $V(\phi)$ is assumed to be Hermitian.
The $\hat{N}+1$ of the last term corresponds roughly to the volume factor $\sqrt{-g}$. 
If this operator was omitted, the potential term would not remain in the continuum limit. 
Note that this potential term breaks the correspondence to the BPS equation \eq{BPS}.
The action, from which the equation of motion \eq{eompotential} can be derived, is given by 
\be
\label{actionscalar}
S_{scalar}=
{\rm Tr}\left(\frac12 
\sum_{i=1}^{2} ([a_i,\phi])^\dagger [a_i,\phi] -   \frac12 \sum_{i=1}^3 ([L_i,\phi])^\dagger 
[L_i,\phi]- (\hat{N}+1) V(\phi) \right),
\ee
where ${\rm Tr}$ is the trace over the whole representation space ${\cal H}=\sum_N \oplus {\cal H}_N$.  

To include gauge symmetry, let me assume that the Hermitian scalar field $\phi$ is also an
operator on an additional representation space ${\cal H}_g$ of a Lie algebra, i.e. 
$\phi=\sum_N \phi_N$, where $\phi_N$ is an operator on ${\cal H}_g\times  {\cal H}_N $.
For simplicity, the Lie algebra and its representation are assumed to be the Lie algebra 
of the unitary group U($n$) and its fundamental representation, respectively. 
Let us consider an infinitesimally small gauge transformation,
\be
\label{scalartrans}
\delta \phi= i \, [g ,\phi],
\ee
where $g$ is an infinitesimally small Hermitian operator in the same class of $\phi$,
\be
\label{sumofg} 
g=\sum_N g_N,
\ee 
where $g_N$ is an operator on ${\cal H}_g \times {\cal H}_N $,

To make the action \eq{actionscalar} invariant under the gauge transformation \eq{scalartrans}, 
let me introduce the gauge fields for $L_i$ and $a_i$,
\bea
\tilde{L}_i&=&L_i+A^L_i,\cr
\tilde{a}_i&=&a_i+A^a_i,
\eea
which transform under the gauge transformation,
\bea
\label{vectortrans}
\delta A^L_i &=& \delta \tilde{L}_i= i\, [g, \tilde{L}_i], \cr
\delta A^a_i &=& \delta \tilde{a}_i=i \, [g, \tilde{a}_i].
\eea
Here the gauge field $A^L_i$ is assumed to be Hermitian.
Because of the property \eq{sumofg}, it can be generally assumed that
the gauge field $A^L_i$ is in the same class as $\phi$, $g$ and $L_i$,
which maps ${\cal H}_g \times {\cal H}_N$ to ${\cal H}_g \times {\cal H}_N$,
while $A^a_i$ is in the same class as $a_i$,  
which maps ${\cal H}_g \times {\cal H}_N$ to ${\cal H}_g \times {\cal H}_{N-1}$.

Let me consider an action, 
\be
\label{actionscalargauge}
\tilde{S}_{scalar}={\rm Tr}\left(\frac12 
\sum_{i=1}^{2} ([\tilde{a}_i,\phi])^\dagger [\tilde{a}_i,\phi]
-\frac12 \sum_{i=1}^3 ([\tilde{L}_i,\phi])^\dagger [\tilde{L}_i,\phi]
-\tilde{N} V(\phi) \right),
\ee
where $\tilde{N}=\frac12 \sum_{i=1}^2 (\tilde{a}^\dagger_i \tilde{a}_i+
\tilde{a}_i \tilde{a}_i^\dagger)$, and the trace is over 
$\sum_N \oplus \ {\cal H}_g \times {\cal H}_N$.
It can be easily seen that the action \eq{actionscalargauge} is invariant under the gauge 
transformation \eq{scalartrans} and \eq{vectortrans},
\be
\delta \tilde{S}_{scalar}={\rm Tr}\left( i\, [g,\tilde{S}_{scalar}]\right)=0.
\ee 

The equation of motion derived from \eq{actionscalargauge} is given by
\be
\label{eomscalargauge}
\sum_{i=1}^3 [\tilde{L}_i,[\tilde{L}_i,\phi]]-\frac12 \sum_{i=1}^{2} 
([\tilde{a}_i^\dagger, [\tilde{a}_i,\phi]+[\tilde{a}_i, [\tilde{a}_i^\dagger,\phi])+
{\rm Sym}\left[\tilde{N} \frac{d V(\phi)}{d \phi}\right]=0,
\ee
where Sym denotes the symmetrized product, 
\be
{\rm Sym}[\tilde{N} \phi^n]=\frac1{n+1}\sum_{i=0}^n \phi^i \tilde{N} \phi^{n-i}.
\ee
The difference of the second and the last terms of \eq{eomscalargauge}
from the expression of \eq{eompotential} is due to the fact that 
$[\tilde{a}_i,\tilde{a}^\dagger_j]\neq \delta_{ij}$ and $[\tilde{N},\phi]\neq 0$ for general $A_i^a$.

To obtain the conserved current coupled with the gauge fields, let us define 
\bea
\label{defcurrentsc}
\tilde{J}_L^i&\equiv&\frac{\delta \tilde{S}_{scalar}}{\delta A^L_i}=-[[\tilde{L}_i,\phi],\phi] , \cr
\tilde{J}_a^i&\equiv&\frac{\delta \tilde{S}_{scalar}}{\delta A^a_i}=
\frac12[[\tilde{a}^\dagger_i,\phi],\phi]
-\frac12 V(\phi)\tilde{a}^\dagger_i-\frac12 \tilde{a}^\dagger_iV(\phi), \cr
\tilde{J}_{a^\dagger}^i&\equiv& (\tilde{J}_a^i)^\dagger.
\eea 
This current satisfies the following conservation law,
\be
\label{conservationlawscalar}
\sum_{i=1}^2 ([\tilde{a}_i,\tilde{J}_a^i]+[\tilde{a}_i^\dagger,\tilde{J}_{a^\dagger}^i])+
\sum_{i=1}^3 [\tilde{L}_i,\tilde{J}_L^i]=0.
\ee
This can be easily proved by using the equation of motion \eq{eomscalargauge}.

In the continuum theory, if a global symmetry exists, conserved charges are associated to 
the symmetry. In the present fuzzy model, if the symmetry associated to ${\cal H}_g$ is 
a global one in place of the gauge symmetry introduced above, 
the following current conservation holds, 
\be
\label{conlawscalglob}
\sum_{i=1}^2 ([a_i,J_a^i]+[a_i^\dagger,J_{a^\dagger}^i])+
\sum_{i=1}^3 [L_i,J_L^i]=0,
\ee 
where $J_L^i,J_a^i,J_{a^\dagger}^i$ are defined by switching off the gauge fields 
in the current \eq{defcurrentsc}. 
To see what are the conserved charges associated to the current conservation 
\eq{conlawscalglob}, let me take the product of \eq{conlawscalglob} and 
an Hermitian generator $T^b$ of the symmetry on ${\cal H}_g$, and take
the trace over $\sum_{N=N_1}^{N_2} \oplus \ {\cal H}_g \times {\cal H}_N$,
\be
\label{conscgl}
{\rm Tr}_{N_1-N_2}\left( 
\sum_{i=1}^2 ([a_i,J_a^{i\ b}]+[a_i^\dagger,J_{a^\dagger}^{i\ b}])
+\sum_{i=1}^3 [L_i,J_L^{i\ b}]\right)= 0,
\ee
where I have used the commutativity between $T^b$ and $a_i,a_i^\dagger,L_i$, and have defined
\bea
J_L^{i\ b} &\equiv& {\rm tr}_g(T^b J_L^i),\cr
J_a^{i\ b} &\equiv& {\rm tr}_g(T^b J_a^i), \cr
J_{a^\dagger}^{i\ b}&\equiv& (J_a^{i\ b})^\dagger,
\eea
where ${\rm tr}_g$ denotes the trace over ${\cal H}_g$.
Because of the property of the trace and that $L_i$ maps ${\cal H}_N$ to ${\cal H}_N$,
the last term of the left-hand side of \eq{conscgl} vanishes,
\be
\label{scchargeeqone}
{\rm Tr}_N([L_i,J_L^{i\ b}])={\rm Tr}_N(L_i J_L^{i\ b})-{\rm Tr}_N(J_L^{i\ b} L_i)
={\rm Tr}_N(L_i J_L^{i\ b})-{\rm Tr}_N(L_i J_L^{i\ b})=0,
\ee
where ${\rm Tr}_N$ denotes the trace over ${\cal H}_N$.
On the other hand, $a_i,J_{a^\dagger}^{i\ b}$ and $a_i^\dagger,J_a^{i\ b} $ 
map ${\cal H}_N$ to ${\cal H}_{N-1}$ and ${\cal H}_{N+1}$, respectively.
Therefore, as for the other terms of the left-hand side of \eq{conscgl}, 
\bea
\label{scchargeeqtwo}
{\rm Tr}_N([a_i,J_a^{i\ b}])&=&{\rm Tr}_N(a_iJ_a^{i\ b})-{\rm Tr}_N(J_a^{i\ b}a_i)=
{\rm Tr}_{N+1}(J_a^{i\ b}a_i)-{\rm Tr}_{N}(J_a^{i\ b} a_i  ), \cr
{\rm Tr}_N([a_i^\dagger,J_{a^\dagger}^{i\ b}])&=&{\rm Tr}_N(a_i^\dagger J_{a^\dagger}^{i\ b})
-{\rm Tr}_N(J_{a^\dagger}^{i\ b}a_i^\dagger)=
{\rm Tr}_N(a_i^\dagger J_{a^\dagger}^{i\ b})-{\rm Tr}_{N+1}(a_i^\dagger J_{a^\dagger}^{i\ b}).
\eea
Using \eq{scchargeeqone} and \eq{scchargeeqtwo}, \eq{conscgl} becomes
\be
-\sum_{i=1}^2 {\rm Tr}_{N_1}(J_a^{i\ b}a_i-a_i^\dagger J_{a^\dagger}^{i\ b})+
\sum_{i=1}^2 {\rm Tr}_{N_2+1}(J_a^{i\ b} a_i - a_i^\dagger J_{a^\dagger}^{i\ b}) =0.
\ee
Thus the conserved charges associated to the symmetry can be defined by
\bea
\label{qnbscalar}
Q_N^b&=&i\,\sum_{i=1}^2 {\rm Tr}_N (J_a^{i\ b} a_i - a_i^\dagger J_{a^\dagger}^{i\ b}) \cr
&=& -i\, \sum_{i=1}^2 {\rm Tr}_{N,g}([T^b,\phi] a_i^\dagger \phi a_i),
\eea
where I have put $i$ in the definition to make the charges real, and have used the explicit
expression for the current.

A gauge invariant action for a non-Hermitian scalar field may be written as 
\be
\label{actioncscalargauge}
\tilde{S}_{cs}={\rm Tr}\left(-\frac12 
\sum_{i=1}^{2} ([\tilde{a}_i^\dagger,\phi^\dagger] [\tilde{a}_i,\phi]+
[\tilde{a}_i,\phi^\dagger] [\tilde{a}_i^\dagger,\phi])
+ \sum_{i=1}^3 [\tilde{L}_i,\phi^\dagger] [\tilde{L}_i,\phi]
-\tilde{N} V(\phi,\phi^\dagger) \right),
\ee
where $\phi^\dagger$ is the Hermitian conjugate of $\phi$, and  $V(\phi,\phi^\dagger)$ is 
assumed to be Hermitian. If the potential is invariant under the global transformation, 
\be
\label{uonesymmetry}
\phi\rightarrow e^{i\theta} \phi,\ \ \phi^\dagger \rightarrow e^{-i\theta} \phi^\dagger, 
\ee
for any real c-number $\theta$, there exists a global $U(1)$ symmetry in the action \eq{actioncscalargauge}.
To see what is the conserved charge associated to the symmetry by the Noether's method, 
let me consider an infinitesimally small shift of the field $\phi$ by 
\bea
\delta \phi&=&i\, \alpha\, \phi, \cr
\delta \phi^\dagger &=& -i \, \phi^\dagger \alpha,
\eea
where $\alpha$ is an infinitesimally small Hermitian operator on ${\cal H}$ 
which maps ${\cal H}_g \times {\cal H}_N$ to ${\cal H}_g \times {\cal H}_N$ for any $N$.
Then the shift of the action is given by
\be
\label{scsshift}
\delta S_{cs}=-i\, {\rm Tr} 
\left( \alpha \left( \sum_{i=1}^2 ([\tilde{a}_i,\tilde{J_c}_a^i]+
[\tilde{a}_i^\dagger,\tilde{J_c}_{a^\dagger}^i])+
\sum_{i=1}^3 [\tilde{L}_i,\tilde{J_c}_L^i] +\phi \frac{\pt}{\pt\phi} \tilde{N}V 
- \left(\frac{\pt}{\pt\phi^\dagger}\tilde{N}V\right)\phi^\dagger  \right) \right), 
\ee
where the current is defined by 
\bea
\label{uonecurrent}
\tilde{J_c}_L^i&=&  -[\tilde{L}_i,\phi]\phi^\dagger+\phi [\tilde{L}_i,\phi^\dagger], \cr
\tilde{J_c}_a^i&=&  -\frac12 \phi [\tilde{a}_i^\dagger,\phi^\dagger]+
\frac12 [\tilde{a}_i^\dagger,\phi] \phi^\dagger, \cr
\tilde{J_c}_{a^\dagger}^i&=& {\tilde{J_c^i}_a}^\dagger,
\eea
and the partial derivatives with respect to $\phi,\phi^\dagger$ denote the partial derivatives of the trace,
\be
\frac{\pt}{\pt \phi} A\equiv\frac{{\rm Tr}(\delta A)}{\delta \phi},\ \ 
\frac{\pt}{\pt \phi^\dagger} A\equiv\frac{{\rm Tr}(\delta A)}{\delta \phi^\dagger}.
\ee  
For example,
\be
\frac{\pt}{\pt \phi} \tilde{N} (\phi^\dagger)^2 \phi^2= 
\frac{{\rm Tr}(\tilde{N}(\phi^\dagger)^2 \delta \phi \phi
+\tilde{N} (\phi^\dagger)^2 \phi \delta\phi )}{\delta\phi}=
\phi \tilde{N} (\phi^\dagger)^2+ \tilde{N} (\phi^\dagger)^2 \phi.
\ee 
From \eq{scsshift}, it is observed that 
the divergence of the current does not vanish for a general potential, 
\be
\label{divergenceuone}
\sum_{i=1}^2 ([\tilde{a}_i,\tilde{J_c}_a^i]+
[\tilde{a}_i^\dagger,\tilde{J_c}_{a^\dagger}^i])+
\sum_{i=1}^3 [\tilde{L}_i,\tilde{J_c}_L^i] =-\phi \frac{\pt}{\pt\phi} \tilde{N}V 
+ \left(\frac{\pt}{\pt\phi^\dagger}\tilde{N}V\right)\phi^\dagger. 
\ee
This equation can be also checked directly by using the equation of motion derived from the action 
\eq{actioncscalargauge}.

Regardless of the fact that the current \eq{uonecurrent} is not divergence free, 
there exists a conserved charge associated to the U(1) symmetry. To see this, let me observe
the right-hand side of \eq{divergenceuone} has the property that
\be
\label{npnp}
{\rm Tr}_{N,g}\left(-\phi \frac{\pt}{\pt\phi} \tilde{N}V + 
\left(\frac{\pt}{\pt\phi^\dagger}\tilde{N}V \right)\phi^\dagger\right)\
={\rm Tr}_{N,g}\left((-n_\phi+n_{\phi^\dagger}) \tilde{N}V \right)
\ee
where $n_\phi$ and $n_{\phi^\dagger}$ are the operators which count the number of 
fields $\phi$ and $\phi^\dagger$ in $V$, respectively.
Because of the U(1) symmetry \eq{uonesymmetry} for $V$, $-n_\phi+n_{\phi^\dagger}$ vanishes in 
\eq{npnp}. Therefore the divergence of the U(1) current \eq{divergenceuone} 
vanishes under the trace ${\rm Tr}_{N,g}$. 
Repeating the same argument as the evaluation of \eq{conscgl}, 
the conserved U(1) charge is obtained as
\bea
Q_N&=&i\,\sum_{i=1}^2 {\rm Tr}_{N,g} \left(\tilde{J_c}_a^i \tilde{a_i}  - \tilde{a_i}^\dagger   
\tilde{J_c}_{a^\dagger}^i\right)\cr
&=& i\, \sum_{i=1}^2 {\rm Tr}_{N,g} \left(\phi^\dagger \tilde{a}^\dagger_i \phi 
\tilde{a}_i-\tilde{a}_i^\dagger \phi^\dagger
\tilde{a}_i \phi \right).
\eea

Finally I would like to discuss momentum conservation.
As is well known in the continuum theory, 
when there exists a killing vector field $\xi_\mu$ preserving the background geometry, 
the vector field $T^{\mu\nu}\xi_\nu$ obtained from a conserved energy-momentum tensor 
$T^{\mu\nu}$ satisfies the current conservation law
\be
\nabla_\mu \ (T^{\mu\nu}\xi_\nu)=0.
\ee
This current defines a conserved charge such as energy and momentum. 
When the background is a sphere changing its size, the killing vector fields are 
the time-independent rotational symmetry, and they define conserved momentums.
In the present case of a fuzzy two-sphere, the SU(2) generators $L_i$ play the role of 
such killing vector fields. As was argued in \cite{Madore:1991bw}, the U($N$) symmetry on
${\cal H}_N$ contained in \eq{scalartrans} is the fuzzy analog of the general 
coordinate transformation on the fuzzy two-sphere. 
Since an energy-momentum tensor is associated to the general coordinate transformation in the continuum
theory, the analog of an energy-momentum tensor in the present model may be given by the conserved
current $J_L^i,J_a^i,J_{a^\dagger}^i$ in \eq{conlawscalglob}.
In fact, by using the current conservation \eq{conlawscalglob} and the explicit form of 
the current, one can easily show the following current-divergence equation,
\be 
\label{curconene}
\sum_{i=1}^2\left( [a_i,L_j J_a^i]+[a^\dagger_i,J_{a^\dagger}^i L_j ] \right)
+\sum_{i=1}^3 [L_i,L_j J_L^i] =[L_j,\cdot]+[\phi,\cdot],
\ee
where $\cdot$ denote the terms the explicit forms of which are irrelevant in the present discussions.
Repeating the above procedure of obtaining conserved charges, the right-hand side of
\eq{curconene} does not contribute and the following conserved charges associated to the 
rotational symmetry are obtained,
\bea
\label{momentum}
Q_N^j&=&i\ {\rm Tr}_N\left(-L_j J_a^i a_i+a_i^\dagger J_{a^\dagger}^i L_j\right) \cr
&=& i\ {\rm Tr}_N \left( [L_j,\phi] \sum_{i=1}^2 a_i^\dagger \phi a_i \right),
\eea
where the explicit form of  $J_L^i,J_a^i,J_{a^\dagger}^i$ and 
${\rm Tr}_N([L_i,\cdot])={\rm Tr}_N([\phi,\cdot])=0$ are used.
The continuum limit of these charges will be shown 
to be the conserved momentums in Section \ref{continuum}. 

\section{Spinor field}
\label{spinor}
The action of a spinor field on a unit two-sphere which is invariant under
the general coordinate transformation, the rotations and the chiral symmetry
is given by \cite{Jayewardena:td}
\be
\label{diraccon}
\int_{S^2} d^2x \sqrt{g} \bar \psi ({\cal D}+i)\psi, 
\ee
where $\psi$ is the two-dimensional spinor and ${\cal D}$ is the derivative part of the Dirac 
operator.
The second term $i$ comes from the non-trivial spin-connection on the unit two-sphere 
and is essential for the chiral symmetry of the action.
The Dirac operators on a fuzzy two-sphere have been derived by several 
authors \cite{Grosse:1994ed,Carow-Watamura:1996wg,Balachandran:2003ay}. 
Their forms depend on the choice of the chiral symmetry on the fuzzy two-sphere. 
The common part which remains in the continuum limit is given by
\be
\label{diractwo}
\sigma^k[L_k,\psi]+\psi,
\ee
where the last term corresponds to the second term in \eq{diraccon}. Here,
comparing with the scalar field, the spinor field
$\psi$ has another index of the SU(2) spinor. 
This spinor index will be suppressed in this section, unless otherwise stated.  
The discussions below are mostly in parallel with the previous section,
so that I will skip the details of the derivations.

A gauge transformation is given by
\be
\delta \psi = i [g,\psi].
\ee
Using the Dirac operator \eq{diractwo}, a gauge-invariant action of a massless
spinor field on an evolving fuzzy two-sphere may be written as
\be
\label{actionferm}
\tilde{S}_{spinor}={\rm Tr}\left( a(\hat N) \left( \sum_{k=1}^3 \psi^\dagger \sigma^k 
[\tilde{L}_k,\psi]+ \psi^\dagger \psi \right) +
\sum_{i=1}^2 i\, ( \psi^\dagger \tilde{a}_i^\dagger \psi \tilde{a}_i -
\tilde{a}_i^\dagger \psi^\dagger \tilde{a}_i \psi) \right),   
\ee
where $a(\hat N)$ is a function of $\hat N$, which is an operator $\hat N \left| v\right>=N
\left| v\right>$ for $^\forall\left| v\right>\in {\cal H}_N$. 
This function $a(\hat N)$ could be determined
by imposing a relation with a scalar field like supersymmetry,
but this will not be discussed in this paper. As is in the previous section,
the $\hat N$ could be gauged into $\tilde N$, 
but this is not needed for the gauge symmetry to hold, because $[g,\hat N]=0$.

In the same way as the previous section, the current conservation associated to 
the gauge symmetry is given by
\be
\label{conservationlawferm}
\sum_{i=1}^2 ([\tilde{a}_i,\tilde{J_f^i}_a]+[\tilde{a}_i^\dagger,\tilde{J_f^i}_{a^\dagger}])+
\sum_{i=1}^3 [\tilde{L}_i,\tilde{J_f^i}_L]=0,
\ee 
where the current is defined by
\bea
\label{defcurrentferm}
\tilde{J_f^i}_L&=&-a(\hat N)\sum_{s,s'} \{ \psi_s, \psi_{s'}^\dagger \sigma^i_{s's}\}, \cr
\tilde{J_f^i}_a&=& \sum_s \left( i \psi_s^\dagger \tilde{a}_i^\dagger \psi_s+
i\psi_s \tilde{a}_i^\dagger \psi_s^\dagger  \right), \cr
\tilde{J_f^i}_{a^\dagger}&=& {\tilde{J_f^i}_a}^\dagger. 
\eea 
Here the summation over the spinor indices is explicitly written.
If the symmetry is a global one, the associated conserved charge is given by 
\be
{Q_f}^b_N=\sum_{i,s=1,2}{\rm Tr}_{N,g}\left(a_i^\dagger\psi_s^\dagger a_i [T^b,\psi_s]
+\psi_s^\dagger a_i^\dagger [T^b,\psi_s] a_i \right).
\ee 

The action \eq{actionferm} is invariant under the global U(1) symmetry, 
\be
\label{fermglobalone}
\psi \rightarrow e^{i \theta} \psi ,\ \ \ \psi^\dagger\rightarrow e^{-i\theta} \psi^\dagger.
\ee
The current conservation associated to the global U(1) symmetry \eq{fermglobalone} is given by
\be
\label{fermdivergenceuone}
\sum_{i=1}^2 ([\tilde{a}_i,\tilde{J_{fc}}_a^i]+
[\tilde{a}_i^\dagger,\tilde{J_{fc}}_{a^\dagger}^i])+
\sum_{i=1}^3 [\tilde{L}_i,\tilde{J_{fc}}_L^i] 
=0,
\ee
where the current is defined by 
\bea
\label{fermuonecurrent}
\tilde{J_{fc}}_L^i&=& -a(\hat N)\sum_{ss'} \psi_{s'} \psi_s^\dagger \sigma^i_{ss'} , \cr
\tilde{J_{fc}}_a^i&=&i\, \sum_s \psi_s \tilde{a}_i^\dagger \psi_s^\dagger , \cr
\tilde{J_{fc}}_{a^\dagger}^i&=&\tilde{{J_f}_a^i}^\dagger.
\eea
The conserved charge associated to the global U(1) symmetry is 
\be
{Q_f}_N=\sum_{i,s=1}^2 {\rm Tr}_{N,g}\left(\tilde{a}_i^\dagger \psi_s^\dagger \tilde{a}_i \psi_s +
 \psi_s^\dagger \tilde{a}_i^\dagger \psi_s \tilde{a}_i\right).
\ee

\section{Gauge field}
\label{gauge}
Let me gather the covariant derivatives into one notation,
\be
(K_1,\cdots,K_7)=(\tilde{L}_i,\tilde{a}_i,\tilde{a}_i^\dagger).
\ee
From the analogy of the continuum theory, the gauge-covariant field strengths may be defined by 
\be
F_{\mu\nu}=i [K_\mu,K_\nu].
\ee
Using these field strengths, a gauge-invariant action may be written as 
\be
\label{actiongauge}
S_{gauge}={\rm Tr}\left(-\frac14 \eta^{\mu\mu'}\eta^{\nu\nu'}F_{\mu\nu}^\dagger F_{\mu'\nu'} - 
\frac12 c_\mu \eta^{\mu\nu} K_\mu^\dagger K_\nu \right),
\ee
where the repeated indices are summed over and the metric $\eta^{\mu\nu}$ 
is a diagonal one,
\be
\label{signaturegauge}
\eta^{\mu\nu}=\left(1,1,1,-\frac{1}{2},-\frac12,-\frac12,-\frac12\right).
\ee 
In \eq{actiongauge}, 
I have added the mass terms of the gauge fields, whose role will be shown below. 

The equations of motion of $A_i^L,A_i^a,A_i^{a^\dagger}$ are given by
\be
\label{eomgauge}
\eta^{\mu\mu'}\left[K_\mu^\dagger, [K_{\mu '},K_\nu]\right]+
\eta^{\mu\mu'}\left[K_\mu, [K_{\mu '}^\dagger,K_\nu]\right] + 2 c_\nu K_\nu=0,
\ee
where the index $\nu$ of the last term is not summed over.
For the background $K_\mu=(L_i,a_i,a_i^\dagger)$ to satisfy the equations of motion, 
the mass parameters must be chosen as 
\bea
c_{1,2,3}&=&-2,\cr
c_{4,5,6,7}&=& -\frac34,
\eea
which are just the values of the Casimir operator $-\sum_{i=1}^3 [L_i,[L_i,\cdot]]$
for $L_i,a_i,a_i^\dagger$. 

This mass term is not multiplied by the factor $N$ like the potential term of the scalar
field theory in Section \ref{scalar}. This mass term is in the order of the size of the sphere 
and can be neglected in the continuum limit of Section \ref{continuum}. 

When the gauge fields are coupled with matters, the total action will be given by adding 
the actions of matters \eq{actionscalargauge}, \eq{actioncscalargauge}, \eq{actionferm} 
to the gauge field action \eq{actiongauge}. Then the equations of motion of the gauge fields will 
be given by adding the currents like \eq{defcurrentsc}, \eq{defcurrentferm} on the right-hand
side of \eq{eomgauge},
\be
\label{eomgaugewithmatters}
\eta^{\mu\mu'}\left[K_\mu^\dagger, [K_{\mu '},K_\nu]\right]+
\eta^{\mu\mu'}\left[K_\mu, [K_{\mu '}^\dagger,K_\nu]\right] + 2 c_\nu K_\nu=
2 \eta_{\nu\nu'}J_{K_{\nu'}^\dagger}.
\ee
The left-hand side of \eq{eomgaugewithmatters} can be easily shown to satisfy the identity,
\be
\eta^{\nu\nu'}\left[K_\nu^\dagger,
\eta^{\mu\mu'}\left[K_\mu^\dagger, [K_{\mu '},K_{\nu'}]\right]+
\eta^{\mu\mu'}\left[K_\mu, [K_{\mu '}^\dagger,K_{\nu'}]\right] + 2 c_{\nu'} K_{\nu'}\right]=0.
\ee
This is consistent with the current conservation \eq{conservationlawscalar} and  
\eq{conservationlawferm}.

\section{Continuum limit and background geometry}
\label{continuum}
In this section I will consider the case that the size of the fuzzy sphere is very large.
It will be shown that, in this case, 
the scalar field theory constructed in Section \ref{scalar} can
be approximated by the continuum scalar field theory on a non-trivial geometric background.
 
Let me expand the scalar field $\phi$ on the fuzzy two-sphere in the following way,
\be
\label{expansionphi}
\phi=\sum_{N,j,m} \phi_{j,m}^N Q_{j,m}^N,
\ee 
where $Q_{j,m}^N\ (j=0,1,\cdots, N;\, m=-j,-j+1,\cdots,j)$ 
are the complete set of the operators on ${\cal H}_N$, and have the properties,
\bea
\label{defofq}
\sum_{i=1}^3 [L_i,[L_i,Q^N_{j,m}]]&=&j(j+1)\, Q_{j,m}^N, \cr
[L_3,Q_{j,m}^N]&=&m\, Q_{j,m}^N,\cr
{\rm Tr}\left( {Q_{j,m}^N}^\dagger Q_{j',m'}^{N'}\right) &=& \delta_{NN'} \delta_{jj'}\delta_{mm'}.
\eea

Let me first discuss the operations $\sum_{i=1}^2 a^\dagger_i \phi a_i$ and 
$\sum_{i=1}^2 a_i \phi a_i^\dagger$ in terms of the expansion \eq{expansionphi}.
Since the operations preserve the SU(2) symmetry and change the spin of the 
representation by one-half, I can assume
\bea
\label{qcoef}
\sum_{i=1}^2 a^\dagger_i Q_{j,m}^N a_i &=& c_{j,m}^{N+} \, Q_{j,m}^{N+1}, \cr
\sum_{i=1}^2 a_i Q_{j,m}^N a_i^\dagger &=& c_{j,m}^{N-} \, Q_{j,m}^{N-1},
\eea
with some coefficients $c_{j,m}^{N+}$ and $c_{j,m}^{N-}$.
To determine the coefficients, let me first consider the first equation of \eq{qcoef}. Then
\be
\label{cbyq}
|c_{j,m}^{N+}|^2=\sum_{l,l'=1}^2{\rm Tr}\left(a_l^\dagger\, Q_{j,m}^N \, a_l \, a_{l'}^\dagger
\, {Q_{j,m}^N}^\dagger \, a_{l'} \right),
\ee
where I have used the normalization condition of \eq{defofq}.
By using an identity for the Pauli matrices,
\be
\delta_{\alpha\dot\alpha} \delta_{\dot\beta\beta}+\sum_{k=1}^3 \sigma^k_{\alpha\dot\alpha} 
\sigma^k_{\dot\beta\beta}=2 \delta_{\alpha\beta}\delta_{\dot\alpha\dot\beta},  
\ee
the right-hand side of \eq{cbyq} becomes
\be
\frac12 (N+2)^2 +2 \sum_{i=1}^3 
{\rm Tr} \left(L_i\, Q_{j,m}^N\, L_i\, {Q_{j,m}^N}^\dagger \right).
\ee
The second term of this expression can be evaluated by  
\bea
2 \sum_{i=1}^3 {\rm Tr} \left(L_i\, Q_{j,m}^N\, L_i\, {Q_{j,m}^N}^\dagger \right)&=&
\sum_{i=1}^3 {\rm Tr} \left( -[L_i,[L_i,Q_{j,m}^N]] {Q_{j,m}^N}^\dagger
+ {L_i}^2\, Q_{j,m}^N\, {Q_{j,m}^N}^\dagger +  Q_{j,m}^N\, {L_i}^2\,{Q_{j,m}^N}^\dagger\right) \cr
&=& \left(-j(j+1)+2 \frac{N}2 \left(\frac{N}2 +1\right)\right).
\eea
Thus, $c_{j,m}^{N+}$ may be chosen as
\be
\label{valuecp}
c_{j,m}^{N+}=\sqrt{(N+1)(N+2)-j(j+1)}.
\ee
In the same way as above, it can be shown that
\be
\label{valuecm}
c_{j,m}^{N-}=\sqrt{N(N+1)-j(j+1)}.
\ee

Using the expansion \eq{expansionphi} and the evolutions \eq{qcoef}, \eq{valuecp}, 
\eq{valuecm}, 
the equation of motion of the massless scalar field \eq{eomscalarone} can be expressed as 
\be
\label{eomphijm}
\left(j(j+1)-2N-2\right)\phi_{j,m}^N+\sqrt{N(N+1)-j(j+1)} \phi_{j,m}^{N-1} +
\sqrt{(N+1)(N+2)-j(j+1)} \phi_{j,m}^{N+1}=0.
\ee 
Since this equation determines the whole values of $\phi_{j,m}^N$ from its values at other
two $N$s unless $j\sim N$, I may assume that the corresponding continuum equation be 
a second order differential equation. Regarding $N$ to be a continuum variable, I substitute 
the Taylor expansion,
\bea
\phi_{j,m}^{N+1}&\approx&\phi_{j,m}^N+{\phi_{j,m}^N}'+\frac12 {\phi_{j,m}^N}'' ,\cr
\phi_{j,m}^{N-1}&\approx&\phi_{j,m}^N-{\phi_{j,m}^N}'+\frac12 {\phi_{j,m}^N}'' ,
\eea
into \eq{eomphijm}, and obtain
\bea
\label{firstappsc}
&&\left(j(j+1)-2N-2+\sqrt{N(N+1)-j(j+1)}+\sqrt{(N+1)(N+2)-j(j+1)} \right) \phi_{j,m}^N \cr
&&+\left(\sqrt{(N+1)(N+2)-j(j+1)}-\sqrt{N(N+1)-j(j+1)}\right) {\phi_{j,m}^N}' \cr
&&+\frac12 \left(\sqrt{(N+1)(N+2)-j(j+1)}+\sqrt{N(N+1)-j(j+1)}\right) {\phi_{j,m}^N}''=0.
\eea
If I further assume 
\be
\label{smallvalue}
\frac{j(j+1)}N\ll 1,
\ee
the second order differential equation \eq{firstappsc} simplifies to 
\be
\label{scalarsimple}
j(j+1)\phi_{j,m}^N + {\phi_{j,m}^N}' + (N+1) {\phi_{j,m}^N}''=0.
\ee
The physical meaning of the condition \eq{smallvalue} will be discussed later. 

On the other hand in the continuum, 
the equation of motion of a free massless scalar field on a non-trivial geometric background 
in three space-time dimensions is given by
\be
\label{eomscalarcont}
\frac{1}{\sqrt{-g}} \pt_\alpha \sqrt{-g} g^{\alpha\beta} \pt_\beta \, \phi =0,
\ee
where $\alpha,\beta=0,1,2$ and the repeated indices are summed over. 
Let me take the gauge that the temporal coordinate is $N$ and assume that the metric has the form,
\be
\label{assumedmetric}
ds^2=- f(N)\, dN^2+ h(N)\, d\Omega^2,
\ee
where $f(N),h(N)$ are the functions of $N$ to be determined, and $d\Omega^2$ is the metric on
the unit two-sphere.
Substituting this metric \eq{assumedmetric} into \eq{eomscalarcont}, the equation of 
motion of a scalar field is
\be
\label{eomscalarcontinuum}
\left( \nabla_\Omega^2 - \frac1{\sqrt{f}}\frac{\pt}{\pt N} \frac{h}{\sqrt{f}}\frac{\pt}{\pt N}
\right) \phi=0,
\ee
where $\nabla_\Omega^2$ is the Laplacian on the unit two-sphere and is known to have the eigenvalues 
$-j(j+1)\ (j=0,1,\cdots)$ with degeneracy $2j+1$. 
Then \eq{scalarsimple} can be regarded as a cut-off 
version\footnote{Note that $j=0,1,\cdots,N$ for the fuzzy two-sphere.} 
of \eq{eomscalarcontinuum}, if
\bea
\frac{1}{\sqrt{f}}\frac{d}{dN} \left(\frac{h}{\sqrt{f}}\right)&=&1,\cr
\frac{h}{f}&=&N+1.
\eea
This determines $f,h$ uniquely \footnote{Up to a trivial overall constant.} as
\bea
f&=&1, \cr
h&=& N+1.
\eea 
Therefore the background metric of the corresponding continuum theory is given by
\be
\label{corrmetric}
ds^2=-dN^2+(N+1)\, d\Omega^2.
\ee

If the scalar field has a potential term, the left-hand side of \eq{eomscalarcontinuum}
contains $-h V'(\phi) = -(N+1) V'(\phi)$. 
This term corresponds to the potential term of \eq{eompotential}.

The solution of the second-order differential equation \eq{scalarsimple} is explicitly given by 
the Bessel function with $2\sqrt{j(j+1) (N+1)}$ as its argument, 
and hence the angular frequency is given by $\sqrt{j(j+1)/(N+1)}$ for large $N$.  
The angular frequency may be identified with the energy associated to the mode. 
Therefore the condition \eq{smallvalue} means that the energies associated to the modes 
must be much smaller than the unit of energy of the present model to justify the 
continuum limit.

The metric \eq{corrmetric} shows that the maximal physical length scale of the space at time $N$
is given by $\sqrt{N}$. This means that the spatial coordinates in the physical length unit 
should be identified with
\be
x_{phys}^i\sim \frac{L_i}{\sqrt{N}},
\ee
because the maximum values of $L_i$ is in the order of $N$. 
To discuss the non-commutativity of the physical coordinates, I may take any location
on the fuzzy two-sphere, say the `north' pole.
Near there, $L_3$ is approximately $N/2$ so that the coordinates tangent to the sphere
have the non-commutativity,
\be
[x_{phys}^1,x_{phys}^2]\sim 1.
\ee
This commutation relation means an area uncertainty, 
\be
\label{minimalarea}
\Delta x_{phys}^1\, \Delta x_{phys}^2 \gtrsim 1.
\ee    
Thus the fuzzy sphere of this model has the minimal area of $O(1)$. 
Since the maximum length scale is given by $\sqrt{N}$, this uncertainty relation
\eq{minimalarea} leads to the minimum length of order $1/\sqrt{N}$. Therefore the maximum
momentum or energy will be in the order of $\sqrt{N}$, which is consistent with the fact
that the maximum energy of the model is $\sqrt{N}$ for $j=N$. 

Let me take the continuum limit of the charge \eq{qnbscalar}. The limits 
of the other charges can be taken in similar ways.
The computation is straightforward as follows.
\bea
\label{qnbcontlimit}
Q_N^b&=&-i\, \sum_{i=1}^2 {\rm Tr}_{N,g} \left( [T^b,\phi]\, a_i^\dagger\, \phi\, a_i \right) \cr
&=& -i \, \sum_{i=1}^2 \sum_{j,m} {\rm Tr}_{N,g} \left( [T^b,\phi^N_{j,m}] \, Q_{j,m}^N \, 
\phi_{j,-m}^{N-1} \, c_{j,-m}^{N-1\, +}\,  Q_{j,-m}^{N}\right) \cr
&=& -i \, \sum_{j,m} c_{j,m}^{N-1\, +}\, {\rm Tr}_g \left([T^b,\phi_{j,m}^N] \phi_{j,-m}^{N-1}
\right) \cr
&\approx&  i\, \left(N+\frac12\right) 
\sum_{j,m} {\rm Tr}_g \left( [T^b,\phi_{j,m}^N] {\phi^N_{j,-m}}'\right),
\eea
where I have used the same approximation as before and have used the expansion  
$\phi_{j,m}^{N-1}\simeq \phi^N_{j,m}-{\phi^N_{j,m}}'$.
This should be compared with the continuum expression 
\be
\label{conscalarcharge}
i \int_N d^2\Omega \, \sqrt{g}\,  {\rm Tr}_g\left( [T^b,\phi]\, \phi'\right)
=i\, h(N)\, \sum_{j,m} {\rm Tr}_g \left( [T^b,\phi_{j,m}] {\phi_{j,-m}}'\right),
\ee
where I have expanded the field $\phi$ in terms of the spherical harmonics which are 
normalized on the unit sphere.
Since the charge $Q_N^b$ depends on the fields at both time $N$ and $N-1$,
the time in \eq{conscalarcharge} should be identified with the mean value $N-\frac12$. 
Then, since $h\left(N-\frac12\right)=N+\frac12$, \eq{qnbcontlimit} and \eq{conscalarcharge} agree
with each other.

To discuss the continuum limit of \eq{momentum}, let me start with 
the energy-momentum tensor of a scalar field with a potential, 
\be
T_{\alpha\beta}=\pt_\alpha \phi \pt_\beta \phi -
g_{\alpha\beta} \left(\frac12 g^{\gamma\delta} \pt_\gamma \phi \pt_\delta \phi +  V(\phi) \right).
\ee
Taking into account the diagonal form of the metric tensor \eq{assumedmetric} or \eq{corrmetric},
the time-component of the conserved current associated to the rotational symmetry is given by
\be
\xi^{(j)}_\alpha \pt^\alpha \phi \pt^0 \phi,
\ee
where $\xi^{(j)}_\alpha$ denote the killing vector fields associated to $L_j$, and have only 
the spatial components. Since $\xi^{(j)}_\alpha \pt^\alpha \phi$ can be identified 
with $[L_j,\phi]$ 
and $\sum_{i=1}^2 a_i^\dagger \phi a_i \approx (N+1/2)(\phi+\pt^0 \phi)$ under the same
approximation as above, 
the continuum limits of \eq{momentum} are actually the conserved momentums of 
the continuum theory.

\section{Summary and discussions}
\label{discussions}
In this paper, I have constructed field theory on the evolving fuzzy two-sphere
proposed in \cite{Sasakura:2003ke},
which is described by a pure-into-mixed-state evolution. 
I have first studied scalar fields. The equation of motion of a massless
scalar field has intimate relation with the equation obtained by dropping the time-derivative
term of the equation of \cite{Banks:1983by}. The construction has been extended to spinor  
and gauge fields.

I have also studied the essential properties for field theory such as action,
gauge invariance, current conservation and conserved charges and momentums. 
Soon after the proposal by Hawking \cite{Hawking:1982dj}, the pure-into-mixed-state evolution 
was criticized because of its difficulty in the charge and energy-momentum conservation
\cite{Gross:mq, Banks:1983by}. Especially in \cite{Banks:1983by}, it was shown that such 
evolutions generally violate locality or energy-momentum conservation. 
On the other hand, the field theories constructed explicitly in this paper have conserved charges 
and momentums, and the equation of motion of the scalar field in the continuum limit of 
Section \ref{continuum} shows no clear violation of locality although there exists non-locality of
fuzziness in the order of a fundamental scale.  
This apparent contradiction of the formulation in this paper to the argument of 
\cite{Gross:mq, Banks:1983by} comes from the difference of the dynamical variables 
which evolve through the pure-into-mixed-state evolution.
The dynamical variables of this paper are the fields themselves, 
while it is a density matrix in the original papers. 
As was discussed in \cite{Sasakura:2003ke}, the pure-into-mixed-state evolutions  
in the Shr\"odinger and the Heisenberg picture are generally distinct.
This is prominent in charge conservation because of the distinct interpretations
of the dynamics. To see this, let me consider a
charge operator $G$ and impose the symmetry $G$ to the superscattering matrix 
$\left| i\right>\left< j\right| \rightarrow {{\$_i}^{jk}}_l \left| k\right>\left<l\right|$.
Then the superscattering matrix satisfies \cite{Gross:mq}
\be
{{\$_i}^{jk}}_l =0 \ \ {\rm unless}\ G_i-G_j=G_k-G_l,
\ee
in a basis which diagonalizes the charge $G$. Therefore, if the superscattering matrix is applied 
to an operator ${\cal O}={C^i}_j \left|i\right>\left<j\right|$, its charge is conserved, namely,
if $[G,{\cal O}]=g\ {\cal O}$, then $[G,\$({\cal O})]=g \ \$({\cal O})$ for an 
eigenvalue $g$.
On the other hand, if it is applied to a density matrix $\rho$, the charge is not generally 
conserved, ${\rm Tr}(\rho G) \neq {\rm Tr}(\$(\rho) G)$ \cite{Gross:mq}.    
Therefore if the evolutions discussed in this paper are described as an evolution of 
a density matrix instead of fields, the dynamical equation for the density matrix 
would take a completely different form from the expression \eq{eachevol}. 
This expectation may circumvent the apparent contradiction 
of the formulation in this paper to the argument of \cite{Gross:mq, Banks:1983by}.

I have taken the continuum limit of the scalar field theory, and have obtained
the background geometry of the corresponding continuum theory. The scale factor of the two-sphere 
is proportional to the square root of the cosmic time.
Interestingly, this behavior of the expansion agrees with the one that saturates the cosmic 
holographic principle proposed in \cite{Fischler:1998st}.
In the model of this paper, this behavior of the scale factor seems to be controlled  
by the mass term of the gauge field, but the origin of this agreement is not clear at present.
This may be made clearer, when the dynamics of the quantum geometry is understood further. 
It would be interesting and challenging to investigate the dynamics of the coupled 
equations of the scalar, spinor and gauge fields, and find more varieties of 
background geometries. 

I have not considered the continuum limits of the spinor and the gauge field theories. 
These limits seem to require more careful identification of the fields on the two-sphere 
than the scalar field. 
Especially there is the question of the exotic signature \eq{signaturegauge}.
The scalar field theory on the evolving fuzzy two-sphere also contains the exotic signature 
in the kinetic term, but this fact does not lead to any complications in the continuum limit, i.e.
there are no extra times, and the continuum limit can be interpreted in the standard way
with a non-trivial background geometry. 
As for the spinor field, the continuum limit will be able to be taken in a more or less similar way as
the scalar field. However, as for the gauge field formulated in Section \ref{gauge}, the dynamical
fields include the extra fluctuations with negative signature which could cause serious 
pathologies. A hint for solving this difficulty can be found in the formulation
of the gauge theory on a fuzzy-two sphere \cite{Carow-Watamura:1998jn}
following Connes' framework of noncommutative geometry \cite{Connes}.
In their formulation it was observed that the action is composed of the standard gauge field
action and an extra term which constrains the gauge field fluctuation transverse to the 
sphere. It might be possible that some extra terms can be added to \eq{actiongauge} to 
kill the unwanted fluctuations with negative signature. 
In this respect, the study of differential noncommutative geometry \cite{Madore:aq}
of the present (2+1)-dimensional fuzzy space-time would be interesting. 

While the continuum limit of the scalar field theory of the present model can be interpreted 
as the standard continuum theory,
deviation certainly exists at high energy or near the birth of the fuzzy space.
Investigating the possibility of detecting the deviation in the present model may give 
suggestions on what can be the signs of the quantum natures of space-time. 
Especially the pure-into-mixed-state evolution of the fuzzy two-sphere is a non-unitary
process, and its effects may pile up to become large enough to be detected 
for a long period of observation.
In fact some uncertainty relations with the property that uncertainties become larger for a 
longer period of observation have been argued by several authors 
\cite{salecker,karolyhazy,Ng:1993jb,Amelino-Camelia:1994vs,Sasakura:1999xp}.
It might be interesting to explore this kind of uncertainty relations in this model.      

\vspace{.5cm}
\noindent
{\large\bf Acknowledgments}\\[.2cm]
The author would like to thank J.~Madore for stimulating discussions during his short stay 
at Yukawa Institute for Theoretical Physics.
The author was supported by the Grant-in-Aid for Scientific Research No.13135213 and No.16540244
from the Ministry of Education, Science, Sports and Culture of Japan.

\end{document}